\documentclass[
 reprint,
superscriptaddress,
 amsmath,amssymb,
 aps,
]{revtex4-2}

\usepackage{graphicx}% Include figure files
\usepackage{dcolumn}% Align table columns on decimal point
\usepackage{bm}% bold math
\usepackage{xcolor}

\begin{document}

\preprint{APS/123-QED}
%\title{Pigment-enhanced structural color}
\title{Efficient structural color from pigment-loaded nanostructures}% Force line breaks with \\

\author{Tianqi Sai}
% \thanks{}
\affiliation{Department of Materials, ETH Z\"{u}rich, 8093 Z\"{u}rich, Switzerland.}

\author{Luis S. Froufe-Pérez}
\affiliation{Department of Physics, University of Fribourg, 1700 Fribourg, Switzerland}

\author{Frank Scheffold}
\affiliation{Department of Physics, University of Fribourg, 1700 Fribourg, Switzerland}

\author{Bodo D. Wilts}
\affiliation{Chemistry and Physics of Materials
University of Salzburg, 5020 Salzburg, Austria}
 
\author{Eric R. Dufresne}
 \email{eric.r.dufresne@cornell.edu}
 \affiliation{Department of Materials, ETH Z\"{u}rich, 8093 Z\"{u}rich, Switzerland.}
 \affiliation{Department of Materials Science and Engineering, Department of Physics, Cornell University, Ithaca, NY, 14850, USA.}

\date{\today}% It is always \today, today,

\begin{abstract}
Color can originate from wavelength-dependence in the absorption of pigments or the scattering of nanostructures. 
While synthetic colors are dominated by the former,  vivid structural colors found in nature have inspired much research on the latter.
However, many of the most vibrant colors in nature involve the interactions of structure and pigment. 
Here, we demonstrate that pigment  can be exploited to efficiently create bright structural color at wavelengths outside its absorption band. 
We created pigment-enhanced Bragg reflectors by sequentially spin-coating layers of poly-vinyl alcohol (PVA) and polystyrene (PS) loaded with $\beta$-carotene (BC). 
With only 10 double layers, we acheived a peak reflectance over $0.8$ at 550 nm and normal incidence.
A pigment-free multilayer made of the same materials would require 25 double layers to achieve the same reflectance. 
Further, pigment loading  suppressed the Bragg reflector's characteristic iridescence.
Using numerical simulations, we further show that similar pigment loadings could significantly expand the gamut of non-iridescent colors addressable by photonic glasses.

\end{abstract}

\maketitle

%\tableofcontents

\section{Introduction}

Structural colors arise from the interference of light scattered from nanostructures \cite{kolle2018}.
Nanostructures with long-range order, such as photonic crystals, can produce bright saturated colors across the full visible spectrum.  
However, these colors tend to vary strongly with viewing angle, \emph{i.e.} they are iridescent \cite{seago2009,kinoshita2005} .
Nanostructures with only short-range order, such as photonic glasses, produce non-iridescent colors. 
However, they tend to be less bright and less saturated,  particularly at longer wavelengths \cite{,magkiriadou2014,kawamura2016, schertel2019a, jacucci2020}.  
In this case, color saturation can be improved by suppressing multiple scattering through the  incorporation of low-concentrations of a broad-band absorber \cite{forster2010}.

Ideally, one could combine the advantages of these two approaches,  and generate structural colors with both low-angle dependence and high saturation.
%We would like to combine the advantages of these two structural color production mechanisms and generate colors with both low-angle dependence and high saturation. 
Structure-pigment interactions can play a key role in achieving this goal. 
Nature’s most vibrant colors often result from a finely tuned interplay between structural and pigmentary color \cite{stavenga2014,henze2019pterin}. 
Some evidence in biological systems suggests that the combined mechanisms can create colors and optical properties unachievable by either mechanism alone \cite{wilts2017extreme,shavit2023tunable}.

In day-to-day life, we create vivid non-iridescent color by starting with a white base and adding pigment.
The white base scatters strongly across wavelengths, and  diffusely reflects most of the light incident upon it.
High-index nanoparticles, such as titanium dioxide or polysytrene, are typically employed as a scattering base.
Scattering is characterised by the transport mean-free path, $\ell_\mathrm{mfp}$, which describes how far light propagates through a material before forgetting its original propagation direction.
Pigment absorbs light over a range of wavelengths.
The absorption length, $\ell_\mathrm{abs}$, characterizes how far a typical photon travels into a material before it is absorbed. 
Wavelengths where $\ell_\mathrm{mfp} \gg \ell_\mathrm{abs}$ are absorbed within the material. 
Wavelengths where $\ell_\mathrm{abs} \gg \ell_\mathrm{mfp}$ are diffusely reflected at high efficiency. 

The contributions of scattering and absorption, however, are not always easy to separate.
They are intimately related, because they both derive from the same material property: the dielectric function. 
The real part of the dielectric function determines the refractive index, which leads to scattering. The imaginary part of the dielectric function determines absorption. 
The real and imaginary parts of the dielectric function cannot be varied independently but must satisfy the Kramers-Kronig relation \cite{lucarini2005}. 
As a consequence, broad-band absorbers, like melanin, can feature an elevated index of refraction ($n \sim 1.7$) across the visible range \cite{wilts2014}.
Melanin nanostructures are widely used in nature for the creation of vivid colors  \cite{sun2013structural}.
Recently, researchers have employed synthetic melanin to produce structural colors \cite{xiao2015,xiao2017,cho2017,heil2023mechanism}. 

Absorbing materials can achieve even higher indices of refraction, but are challenging to use for structural coloration.
Broad-band absorbers with higher refractive indices are inappropriate for structural color applications because $\ell_\mathrm{abs}$ becomes too small to enable constructive interference.
Consider familiar amorphous carbon.  
It absorbs strongly across the visible spectrum and features refractive indices approaching 2.5 \cite{arakawa1977optical}.  However, its absorption length is much shorter than the wavelength of light, making constructive interference impossible.
Semiconductors are a more suitable choice.  
Titanium dioxide has low absorption across the visible spectrum, but absorbs strongly at shorter wavelengths.
Thanks to the Kramers-Kronig relation, it features a refractive index exceeding 2.5 across the visible spectrum \cite{devore1951refractive}. 
From an optics perspective, this is a perfect candidate for structural color.
From a materials perspective, however, there are no scalable methods currently available for creating photonic nanostructures from this material.
Nevertheless, titanium dioxide remains a excellent example of how absorption over one range of wavelengths can increase the refractive index at wavelengths outside the absorption band.

To fuse processibility and high refractive index, absorbing polymers are a promising choice.
UV-absorbing pigments are frequently added to increase the refractive index in the visible spectrum, and provide a  modest increase of the refractive index ($\Delta n \approx 0.07$) \cite{hanemann2011tailoring,hanemann2014viscosity,magrini2019transparent}.
Recently, very high refractive indices in the visible were achieved by blending an everyday polymer, polystrene, with a common plant-based pigment, beta-carotene \cite{yasir2021}.
In that case, refractive indices as high as 2.2 were achieved near the absorption edge (about 525 nm) and decayed slowly to 1.75 at 700 nm.  

In this work, we demonstrate how pigment-enhanced scattering in a processable polymer can be exploited to achieve high-reflectance structural colors. Through a combination of experiment and simulation, we demonstrate how engineered absorption can simplify the production of Bragg mirrors and remove iridescence while maintaining brightness and saturation. 
Using numerical simulation, we show that the interplay of absorption and scattering can expand the gamut of photonic glasses.

\section{Results and Discussion}

\subsection{Optical properties of BC/PS blends}

\begin{figure}
    \centering
    \includegraphics{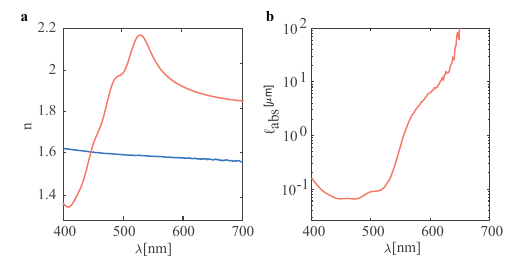}
    \caption{\emph{Effect on beta-carotene (BC) loading on the optical properties of polystyrene (PS).} a) Measured refractive index, $n$, of pure PS (blue) and  $40\%$BC/PS (orange)  b)  Measured absorption length, $\ell_\mathrm{abs}$) of $40\%$BC/PS.}
    \label{fig:nk}
\end{figure}

Following \cite{yasir2021},  we prepared thin composite films of polystyrene (PS) loaded with beta-carotene (BC).
Their optical properties were characterized by ellipsometry.
Results for a PS film loaded with 40\% BC, by weight, are shown in Fig. \ref{fig:nk}.
The blend absorbs strongly at wavelengths up to about 525 nm.  
The refractive index peaks to 2.15 at 528 nm, and decays slowly to 1.8 at 700 nm.
Unfortunately, the absorption length is only 150 nm at the peak refractive index.
This precludes use for structurally colored materials, where light needs to constructively interfere across multiple reflections.
However, with a small shift of wavelength off the peak (to 550 nm), the refractive index enhancement remains highly elevated ($n=2.03$), but the absorption length increases by a factor of 7 to 1.12 $\mu$m.
While this is too short for applications involving bulk propagation, we will show that it can be effectively exploited for nanophotonic applications. %\TS{numbers changed}

\begin{figure*}
    \centering
    \includegraphics{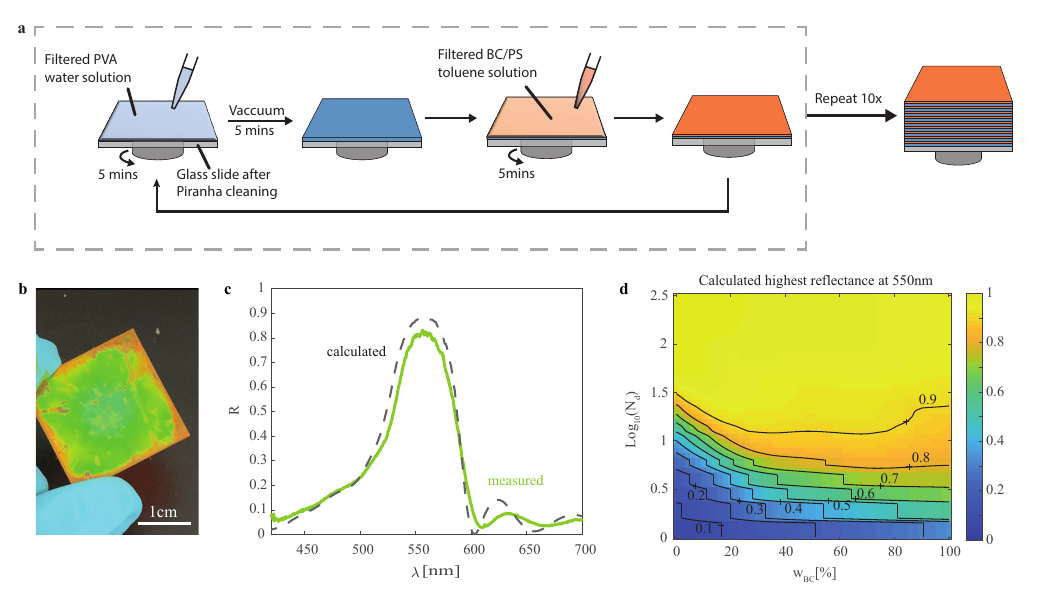}
    \caption{\emph{Enhanced reflectance of pigment-loaded multilayered reflectors.} a) Schematic diagram of the preparation procedure of a 10-double layer reflector made of alternating 40\%BC/PS and PVA films. b) Photo of the prepared reflector.  c) Reflectance of a 10-double layer Bragg reflector, consisting of alternating 40\%BC/PS and PVA films, at normal incidence. Each 40\%BC/PS layer is $57\,\mathrm{nm}$ thick and each PVA layer is $112\,\mathrm{nm}$ thick. The substrate is a plain glass slide. Average measured result of three different point is shown as a green solid line.  The prediction from T-matrix calculations is shown as a dark gray dashed line. d) Colormap of calculated peak reflectance at the wavelength of $550\,\mathrm{nm}$ for multilayers with different number of repeating doublelayers, $N_d$, and different BC concentrations in PS, $w_{BC}$. The optical properties at arbitary concentrations were determined by linear interpolation of the ellipsometry data in \cite{yasir2021}.}
    \label{fig:R}
\end{figure*}

\subsection{Design and Fabrication of a pigment-enhanced Bragg Reflector}

To demonstrate the feasibility of pigment-loaded polymers for structural color, we created Bragg reflectors based on BC-loaded PS blends. 
Bragg reflectors are periodic stacks of alternating dielectric layers whose spacing is optimized to constructively reinforce reflected light over a specific range of wavelengths.
We fabricated Bragg reflectors using a sequential spin coating method \cite{kolle2011photonic,castillo2022enhanced}, shown schematically in Fig. \ref{fig:R}.
In this approach, the solvents of the alternating polymers must be mutually orthogonal. 
At the same time,  each solution should have reasonable wettablity on films of the previous polymer. Commonly, UV ozone or plasma treatment is used between the spinning of each layer to increase the wettablity and to allow the successful deposition of the next layer \cite{wei2018,kasanen2009}. However, BC is degraded by these treatments. 
After exploring a wide range of solvent and polymer combinations, we settled on BC/PS in toluene and polyvinyl alcohol (PVA, $n=1.48$ at 550 nm) in water \cite{castillo2022enhanced}.

The  thicknesses of the polymer layers were optimized using transfer matrix simulations \cite{born2013}, implemented in MATLAB \cite{pascoe2001}. 
To find the optimal layer thicknesses, we fixed the materials to PVA and  $40\%$ BC/PS, and the number of double-layers to 10. 
We varied the thickness of each double layer, $d$, from $120\,\text{nm}$ to $250\,\text{nm}$ and the thickness of the high index layer, $t_{h}$ from $0$ to $d$.
 We   determined the thickness combination which gave the highest reflectance for each wavelength from $400$ to 700 nm. 
 Outside the absorption band, the optimal thickness followed the classic quarter-wave design rule \cite{born2013,schenk2013japanese}, as shown in Figure \ref{fig:ml_wl} a-c.
 The highest reflectivity, about 80\%, was found at  $550\,\text{nm}$, near the edge of the absorption band (Figure \ref{fig:ml_wl}d)
 There, 40\% BC/PS features a refractive index of  $n=2.03$ and absorption length of $\ell_\mathrm{abs}=1.12~\mathrm{\mu}$m. 
% \ED{SAI TO CHECK OR ERIC TO INTERPOLATE FROM THE DATA IN FOLDER}

\subsection{Characterization and Analysis of a pigment-enhanced Bragg Reflector}

To realize this design, we fabricated a 10 double-layer reflector made of of alternating 57 nm thick 40\% BC/PS and 112 nm thick PVA films.  
The cm-scale film had a vivid green appearance, as shown in Fig. \ref{fig:R}b.
Specular reflectance gave the film a mirror-like appearance,  shown in the Movie S1.
We measured the spectrum of reflected light at normal incidence, shown as a solid line in Fig. \ref{fig:R}c. 
The spectrum had a peak reflectance of 83\% in a band near 550 nm, and showed excellent agreement with the spectrum predicted from T-matrix calculations (dashed line in Fig. \ref{fig:R}c). 

To explore the benefits and limitations of pigment-enhanced Bragg mirrors, we used the T-matrix approach to calculate the the reflectance for a range of Bragg reflectors. 
%The optical properties were measured at seven different  BC concentrations (0\%, 10\%, 33\%, 40\%, 50\%, 67\% and 100\%). The optical properties at other concentrations were determined by linear interpolation. 
Fig. \ref{fig:R}d shows the peak reflectance at 550 nm for BC loadings, $w_{BC}$, from 0 to 100\%, and numbers of double layers, $N_d$, from 1 to 316.
Black lines show contours of constant reflectance. 
At low pigment concentrations, $N_d$ decreases exponentially with $w_{BC}$.  
However, the effect of added pigment saturates, particularly for high reflectance.
For a target reflectance of 90\%, the required number of  double layers, $N_d$, drops from 32 to 13 double layers as the BC loading reaches $27\%$.  Additional BC offers no further improvement.
%At 90\% reflectance, $N_d$ is reduced from 32 to 13 double layers as the BC loading reaches $27\%$.  Additional BC offers no further improvement.
At 80\% reflectivity, $N_d$ is reduced by more than a factor of three, from 25 to 8 double layers, at $w_{BC}=40\%$.  Again, additional pigment loading offers little benefit.
The benefit of pigment-loading is strongest as the refractive-index difference  between the undoped layers vanishes (Figure \ref{fig:ml_deltan}) and for wavelengths at the edge of the absorption band (Figure \ref{fig:sup_edge}).

\begin{figure}
    \centering
    \includegraphics[width=0.5\textwidth]{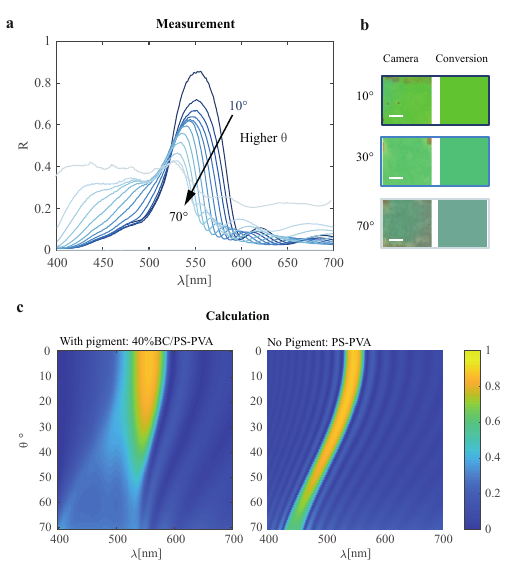}
    \caption{\emph{Reduced iridescence of pigment-loaded Bragg reflector.} a) Measured specular reflectance at different angles of incidence, from $10^{\circ}$ to $70^{\circ}$ with an interval of $5^{\circ}$, of a 10-doublelayer Bragg reflector, consisting of alternating 40\%BC/PS ($57\,\mathrm{nm}$) and PVA ($112\,\mathrm{nm}$) films. b) Photos of the Bragg reflector taken at different angles, and the corresponding RGB color swatches determined from the reflectance spectrum. Scale bar: 1mm. c) T-matrix predication of the  specular reflectance spectra at different angles of incidence.  On the left, results for a 10-doublelayer Bragg reflector, consisting of alternating 40\%BC/PS ($57\,\mathrm{nm}$) and PVA ($112\,\mathrm{nm}$) films. On the right, results for  a 25-double layer Bragg reflector, consisting of alternating PS ($87\,\mathrm{nm}$) and PVA ($90\,\mathrm{nm}$) films. The thicknesses combinations are chosen to give the highest reflectance at the wavelength of $550\,\mathrm{nm}$ }
    \label{fig:angle}
\end{figure}

Pigment-loaded Bragg reflectors not only require fewer double layers, but also show reduced iridescence.
The specular reflectance of the film  measured at different incident angles is shown in Figure \ref{fig:angle}a.
With higher angles of incidence, the reflectance peak vanished, while shifting slightly from $550\,\mathrm{nm}$ to $525\,\mathrm{nm}$.
Images of the film viewed in specular reflection are shown in Figure \ref{fig:angle}b.
At low angles (close to normal incidence), the film shows bright and saturated greens. 
At higher angles, the color is dulled.
There is good agreement between the measured reflectance spectra and the observed color.  To the right of each photo, we show the RGB color inferred from the measured reflectance spectrum using 
the  open source python package \emph{colorpy}.
These results are in good agreement with T-matrix simulations, shown on the left of Figure \ref{fig:angle}c. For comparison, the calculated reflectance spectra from a conventional Bragg mirror (PS-PVA without pigment) are shown in the right panel of Figure \ref{fig:angle}c. This pigment-free reference was designed to give the same peak reflectance at the wavelength of $550\,\mathrm{nm}$ and normal incidence.
In the pigment-free case, we observe a strong shift to shorter wavelengths in the reflectance at larger angles of incidence.  This iridescence is a characteristic feature of conventional Bragg reflectors. In the pigment loaded mirror, iridescence is suppressed because the peak wavelength is shifted into the absorption band of the pigment. 

\subsection{Pigment-enhanced scattering could extend the gamut of photonic glasses}

Having established the viability of pigment-enhanced structural color in Bragg mirrors, we used finite-difference-time-domain simulations to explore its potential in photonic glasses.
Photonic glasses are excellent contrasts to Bragg mirrors.
While the former are 3-D and feature only short-range correlations, the latter are 1-D and have perfect long-range order.
Photonic glasses are made from jammed packings of hard spheres using the protocol described in \cite{skoge2006packing}.

\begin{figure*}
    \centering
    \includegraphics[width=\textwidth]{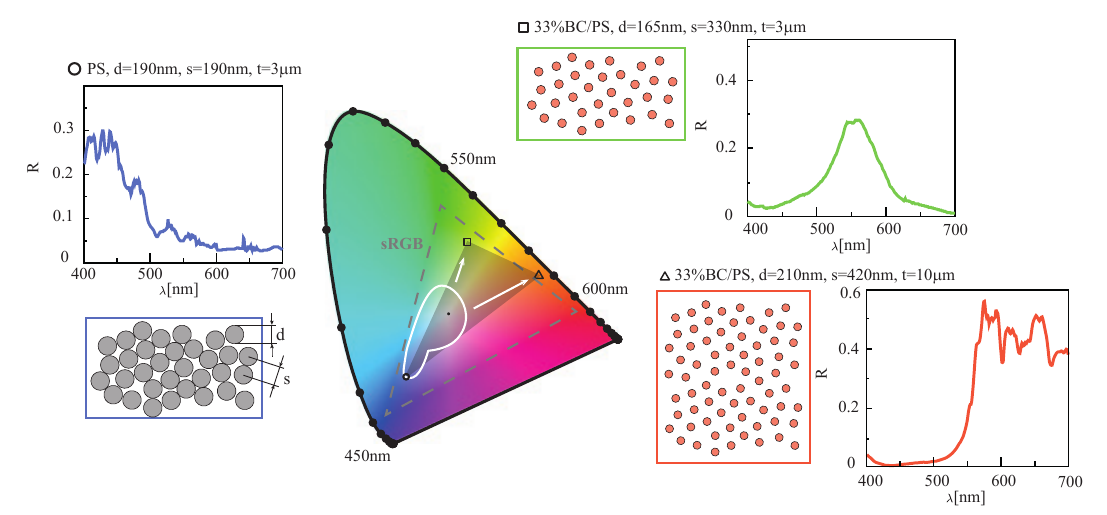}
    \caption{\emph{Extending the gamut of photonic glasses with the help of pigments.} In the center is the CIE chromaticity diagram, with the sRGB gamut indicated by a gray dashed line.  The white line outlines the simulated gamut achieved by amorphous assemblies of pigment-free polystyrene particles. The most saturated blue is marked as a white circle. With the help of BC, simulations of photonic glasses with varying size, spacking and thickness revealed  relatively saturated green (square) and orange-red (triangle) hues. The geometries and spectra of the corresponding photonic glasses are demonstrated in schematics and plots close by. Green (square):$33\%$BC/PS composite particle assembly with a film thickness of $3\mu m$, a particle diameter of $165\,\mathrm{nm}$ and a particle spacing of $330\,\mathrm{nm}$ (square). Red (triangle): $33\%$BC/PS composite particle assembly with a film thickness of $10\mu m$, a particle diameter of $210\,\mathrm{nm}$ and a particle spacing of $420\,\mathrm{nm}$. Blue (circle): PS particle assembly with a film thickness of $3\mu m$, a particle diameter of $190\,\mathrm{nm}$ and a particle spacing of $190\,\mathrm{nm}$.}
    \label{fig:pg_new}
\end{figure*}

Our  explorations suggest that pigment-loading could significantly expand the palette of photonic glasses.  
Without absorption, the gamut from plain polystyrene assemblies is outlined on the CIE colorspace with a white line in Figure \ref{fig:pg_new}. The results of individual simulations, with varying particle spacing and volume fraction, are shown in Figure \ref{fig:fullcp}. This gamut is largely limited to the unsaturated center of the diagram, but accesses some relatively   saturated blues. 
With a  loading of $33\%$ BC inside the PS particles, the gamut can be remarkably extended, as indicated by the gray triangle in the CIE diagram of Fig. \ref{fig:pg_new}. 
The most highly saturated red, green, and blue we were able to create are found at the corners of the triangle. 
This extended gamut fills a large fraction of  with the sRGB gamut (shown as a grey dashed triangle), the standard color space for the internet \cite{anderson1996proposal}. 

The function of the pigment is markedly different in the saturated green and red samples. 
For the green sample, the color emerges thanks to the enhanced refractive index of the BC/PS blend in the green.  This leads to enhanced scattering and structural color.  For the red sample, the color emerges from broadband scattering and the absorption of BC/PS at short wavelengths, similar to a conventional paint.
While first mechanism is sensitive to the thickness of the sample, the second is not, see Table S4.

\section{Conclusion and Outlook}

Our results show that the pigment can enhance the efficiency and gamut of structurally-colored materials.  %can potentially fill in the gaps that traditional structural color mechanisms have. 
For Bragg reflectors, pigment can create high-brightness colors with reduced angle-dependence and less material. For photonic glasses, the same pigment can be used to create one color through pigment-enhanced scattering, or another though conventional absorption in a broad-band scattering medium.  

%\ED{COMPARE AND CONTRAST WITH MELANIN AND ARTIFICICIAL MELANIN - MORE FLEXIBLE PROCESSING?  DETERMINED BY HOST? COMPARE INDICES}

While the  assembly of a pigment-loaded Bragg mirror \emph{via} sequential spin-coating was sufficient for proof-of-principle, the integration of pigments into self-assembled Bragg mirrors \cite{sveinbjornsson2012rapid} would simplify processing.
Several challenges in materials chemistry and processing must be overcome to realize the photonic glasses proposed in this work. 
For example, new synthetic approaches are required to produce monodisperse polymer nanoparticles with high pigment loadings. 
In the meantime, a systematic numerical exploration of the design space would more clearly define the benefits and limitations of this approach.
%Further

\section{Ackowledgements}
We acknowledge the Swiss National Science Foundation, National Centre of Competence in Research ‘Bio-Inspired Materials’ for funding, as well as Miguel Castillo, Ullrich Steiner, and Bodo Wilts for helpful discussions.

\section{Materials and Methods}
\subsection{Materials}
BC (synthetic, $\geq$93\% (UV)), PS (35 kDa (used for making films and multilayers) and 192 kDa (used for processability experiments)), PVA (Mw 85,000-124,000, 99+$\%$ hydrolyzed), toluene and dichloromethane were bought from Sigma-Aldrich and used as received.

\subsection{Spin Coating}

For the preparation of the PVA aqueous solution($2\%$), $2\,\text{g}$ PVA powder was added into 100 mL of milli-Q water. Once added, the solution was stirred vigorously on a hotplate at 95 °C for one hour to obtain a homogenous solution. For the preparation of the $40\,\%$BC/PS toluene solution ($10\,$mg per mL), $60\,\text{mg}$ BC, $90\,\text{mg}$ PS and $15\,\text{mL}$ toluene were added to a $15\,\text{mL}$ glass vial (to minimize the amount of air left in the vial). Then, the vial was sealed, and the solution was stirred at $800\,\text{rpm}$ for $30\,$ minutes. The glass vial was covered with aluminum foil during stirring to avoid light. After stirring, the solution was filtered using a syringe filter (pore size $0.45\,\mu m$).

The glass slide ($2.5\,\text{cm}\times2.5\,\text{cm}$) for spin coating was used after piranha cleaning. For the deposition of multilayers, $1\,\text{mL}$ PVA solution was first spread on the glass slide and spun at $1500\,\text{rpm}$ for $5\,\text{mins}$.
After spinning, the film was vacuumed in a vacuum chamber for at least 5 minutes to dry out the water completely before the deposition of the next BC/PS layer. To deposit the BC/PS layer, $0.5\,\text{mL}$ BC/PS toluene solution was spread on the dried PVA layer and spun at $2000\,\text{rpm}$ for $5\,\text{mins}$.
The same procedure was repeated 10 times for the preparation of every double layer, as shown in Figure \ref{fig:R}a. The spectrum of the prepared film was characterized freshly right after preparation.

\subsection{Ellipsometry}

 A spectroscopic ellipsometer (M2000, J.A. Woollam) was used to measure the phase and amplitude change of light after interacting with the samples.
All measurements were performed between 250 and $1000~\mathrm{nm}$ at an angle of incidence $65^{\circ}$, and all data were acquired and analyzed using WVASE software. Since the films are transparent at longer wavelengths, the films were regarded as a homogeneous material with the thickness fitted by Cauchy dispersion relation in the wavelength range from 600 to $1000~\mathrm{nm}$, After the film thickness was determined and fixed, the optical constants, refractive indices $n$ and extinction coefficient $\kappa$ were fitted using point to point fitting mode.

\subsection{Reflectance measurement}

The reflectance at normal incidence of the prepared Bragg reflector was measured on a microscope (Nikon Ti2) with a 60x air objective (S Plan Fluor ELWD, NA=0.7). The light source is a broadband halogen fiber-coupled illuminator (Thorlabs OSL2). The field aperture and the angular aperture on the microscope were closed to the minimum to reduce the spot size and cut off higher angle reflectance. The final spot size was around $50 \mu m$, and the collecting angle was about $15^{\circ}$. The reflected light was collected by a light detector (Ocean Optics, Ocean HDX). The reflectance was measured at three different points and averaged.

The specular reflectance of the films have been measured with an angle-resolved spectrophotometry setup. The light source is a deuterium-halogen lamp (Ocean Optics DH-2000-BAL), ranging from $200~\mathrm{nm}$ to $1000~\mathrm{nm}$. The sample is placed on a rotating stage and can be tilted to the desired angles. Finally, a light detector (Ocean Optics, QE Pro) is mounted on a rotating arm, which is controlled by software. The incident angle varied from $10^{\circ}$ to $70^{\circ}$. 

For both setups absolute reflectance is calculated using reflection from a broadband dielectric mirror Thorlabs, BBSQ2-E02) at the same incident angle. During the reflectance measurements, a layer of translucent Scotch tape was applied on the back side of the glass substrate to suppress unwanted back reflection from the glass-air interface.

\subsection{FDTD simulations}
In this work, the optical properties of 3D photonic glasses of strongly correlated hard spheres were calculated.
Maximally random jammed assemblies were generated using the algorith presented in \cite{skoge2006packing}. A collection of Poisson-distributed points is taken as an initial configuration in a periodic box. Points are grown in diameter at a certain rate, while collisions are taken into account to avoid overlap until the desired packing fraction is reached. The samples were generated using  the C++ sphere packing code available in https://torquato.princeton.edu/links-and-codes/ (available upon registration). 
In the examples shown, a packing fraction of 64.5\% is reached.
Each simulation considers the optical responses of 1000-3000 of spheres. 

The optical properties of the generated structures were then calculated using LUMERICAL, a commercial-grade software using the FDTD method. Periodic boundary conditions in the lateral Y and Z direction, which are perpendicular to the incoming beam, and perfect matching layer boundaries in the X direction were used in all of the calculations. The incoming source was set as a plane wave.

\nocite{*}

%\bibliography{ref}% Produces the bibliography via BibTeX.
\newpage

%&\ &

\newpage

\setcounter{figure}{0}
\renewcommand{\thefigure}{S\arabic{figure}}%

\appendix

\section{Supplemental Information}

\subsection{Design rule for a pigment-loaded multilayer reflector}

\begin{figure}
    \centering
    \includegraphics[width=0.5\textwidth]{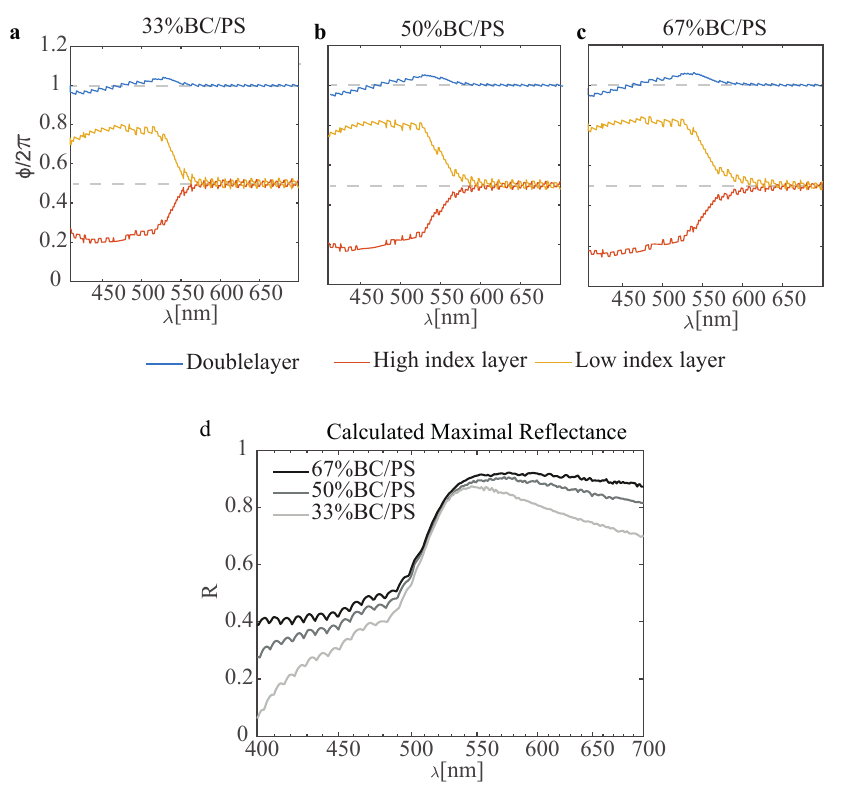}
     \caption{\emph{Design rule for a pigment-loaded multilayer reflector.} (a)-(c) Calculated phase difference after passing through a doublelayer, a high index layer (33w/w\% BC/PS, 50w/w\% BC/PS and 67w/w\% BC/PS respectively) and a low index layer (PVA) with optimal thicknesses. $\phi=\pi$ and $\phi=2\pi$ are marked in dashed grey lines in the plots as guidance. (a) Calculated maximal reflectances, consisting of $10$ doublelayers of alternating low refractive index (PVA) and high refractive index films (33w/w\% BC/PS, 50w/w\% BC/PS and 67w/w\% BC/PS respectively) with optimal thicknesses. The substrates are infinite slabs of glass.}
    \label{fig:ml_wl}
\end{figure}

According to the classical design rule for a quarter-wave mirror, the phase changes after passing through a double layer, a high index layer, and a low index layer are $4\pi n_{d}d/\lambda=2\pi$, $4\pi n_{h}t_{h}/\lambda=\pi$ and $4\pi n_{l}t_{l}/\lambda=\pi$, respectively. $d$, $t_{h}$, $t_{l}$ are the thickness, and $n_{d}$, $n_{h}$, $n_{l}$ are the (average) refractive index, of a double layer, a high index layer, and a low index layer, respectively. To check whether the classical design rule is still valid in the presence of absorption and strong dispersion, we calculate the phase changes, $\phi$ for multilayers with calculated optimal thickness combinations (which gives the highest reflectance), and plot them in Figure \ref{fig:ml_wl} a-c. They are $4\pi n_{d}d/\lambda$ and $4\pi n_{h}t_{h}/\lambda$ and $4\pi n_{l}t_{l}/\lambda$ for a doublelayer, a high index layer and a low index layer, respectively.
It is shown that when outside the absorption band, after $552\,\text{nm}$, $560\,\text{nm}$, $570\,\text{nm}$ for 33w/w\% BC/PS, 50w/w\% BC/PS and 67w/w\% BC/PS, the optimization agrees with the classical design rule.

The optimal peak reflectances are plotted in Figure \ref{fig:ml_wl} a.
It is shown that for all three BC concentrations, the optimal reflectances are the highest in the range around $550\,\text{nm}$, where the high refractive index and low absorption window as expected. The reflectance can be as high as $0.89$ at the wavelength of $573\,\text{nm}$ for 67w/w\% BC/PS. 
From $400\,\text{nm}$ to $500\,\text{nm}$, the reflectance is around $50\%$ because of the large absorption, which is the imaginary part of the complex refractive index. 
For all three concentrations, the optimal reflectance drops slowly towards higher wavelengths because the refractive index enhancement gets weaker.

\subsection{Role of Initial Refractive Index Contrast}
We have discussed how the BC concentration and the number of repeating layers influence the reflectance of a Bragg reflector, assuming it consists of alternating PVA and BC/PS films. The refractive index of PVA is about $0.1$ lower than that of PS, and the initial refractive index contrast when the BC concentration is zero, is 0.1.

\begin{figure}
    \includegraphics[width = \linewidth]{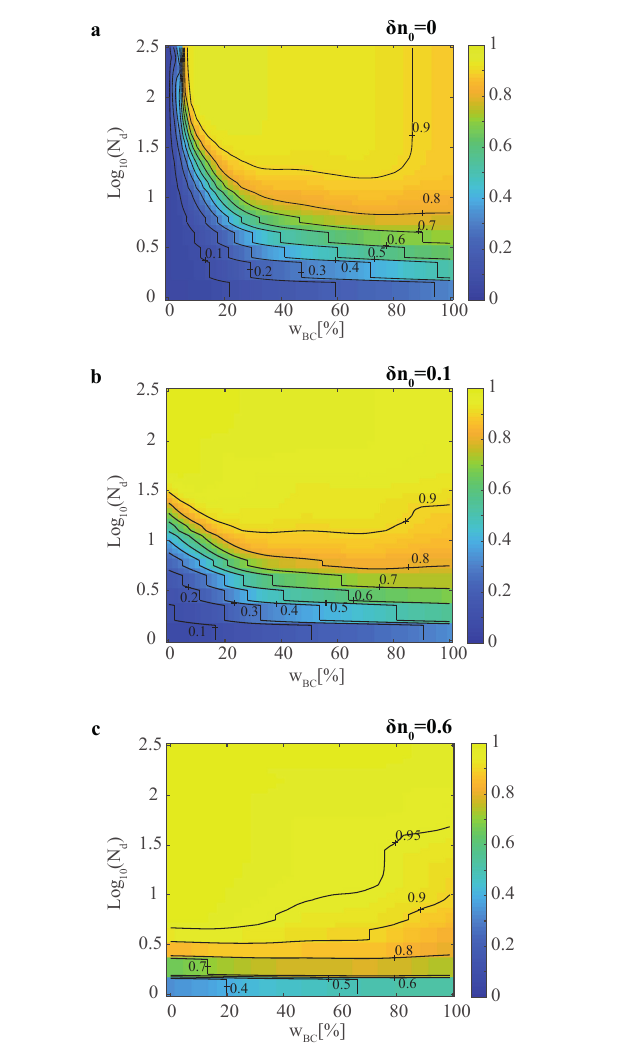}
    \caption{\emph{Effect of initial refractive index contrast.} a)-c) Colormap of calculated maximal reflectance at the wavelength of $550\,\text{nm}$ for multilayers with different numbers of repeating doublelayers and different BC concentrations as high refractive index layers, with $\delta n_{0}$ equals to 0, 0.1 and 0.6 respectively. }
    \label{fig:ml_deltan}
\end{figure}

When there is no initial refractive index contrast, and all the refractive index contrast comes from the incorporation of BC, the BC concentration shows a much stronger effect on the reflectance, as shown in Figure \ref{fig:ml_deltan} a. We observe that with low BC pigment loadings (below $5\%$), the refractive index contrast is too low to give a fair reflectance even when there are hundreds of double layers. By adding more BC ($20\%$-$30\%$), the reflectance would increase significantly. The contour lines in Figure \ref{fig:ml_deltan} a are much steeper than in Figure \ref{fig:ml_deltan} b when BC concentration is below $30\%$.  

If we make the initial refractive index contrast even larger (for example, $\delta n_{0} = 0.6$, which is the refractive index contrast between PS and air), high BC concentration would always make the reflectance lower, as shown in Figure \ref{fig:ml_deltan} c.  

In conclusion, absorption is beneficial in enhancing the multilayer reflectance when the initial refractive index contrast is low. It has a negative effect when there is sufficient initial contrast. 

\begin{figure*}[htbp]
\centering
\makebox[\textwidth]{%
\includegraphics[width=16cm]{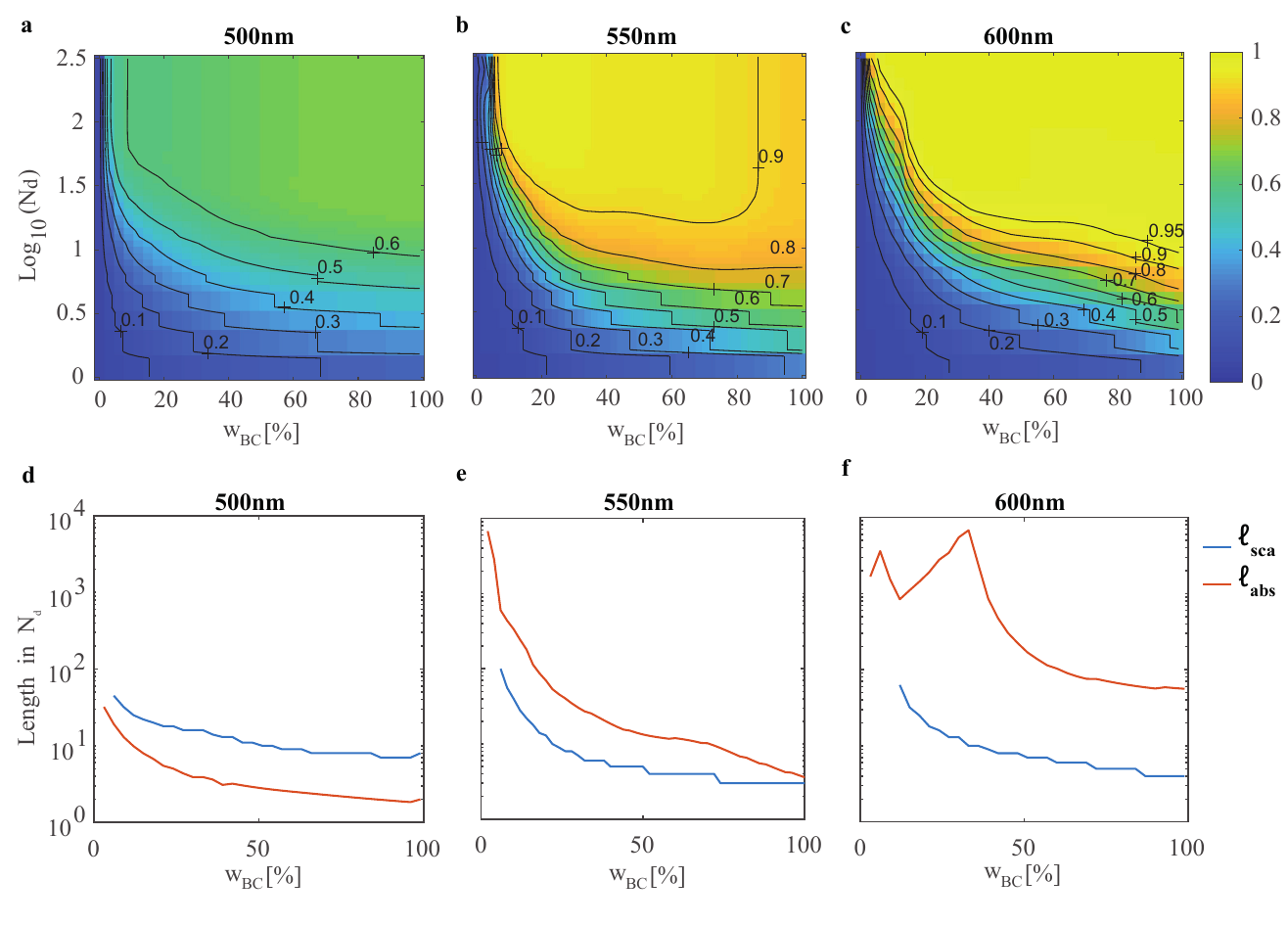}%
}
\caption{\emph{Competition between two length scales.} (a)-(c) Colormaps of calculated maximal reflectance for multilayers with different numbers of repeating doublelayers and different BC concentrations as high refractive index layers with $\delta n_{0} = 0$, at the wavelength of $500\,\text{nm}$, $550\,\text{nm}$ and $600\,\text{nm}$, respectively. (d)-(f) Corresponding absorption lengths and scattering lengths in number of repeating doublelayers at the wavelength of $500\,\text{nm}$, $550\,\text{nm}$ and $600\,\text{nm}$, respectively. }\label{fig:ml_l}
\label{fig:sup_edge}
\end{figure*}

\subsection{Competition of Absorption and Scattering in a Bragg Refelector}

To obtain a complete understanding of the competition between absorption and scattering, we conduct the same optimization and find the highest reflectivities for varying BC concentrations and numbers of doublelayers at $500\,\text{nm}$ and $600\,\text{nm}$ as well, plotted in Figure \ref{fig:ml_l} a-c. 
We also compare the absorption length, $\ell_{abs}$ and the scattering length, $\ell_{sca}$ in Figure \ref{fig:ml_l} d-f. Here, the scattering length is taken as the number of doublelayers needed to reflect back most of light (R is larger than 0.63) without absorption, but with the change of refractive index caused by absorption. This means that we input the true real refractive index of BC/PS composites and set the imaginary part to zero. This can not be true in reality because real and imaginary part of the refractive index come as a ``package''. However, separating the real and imaginary parts here helps us to disentangle the complex problem and understand the underlying physics.

The comparison of absorption length and scattering length helps us understand the reflectance colormaps.
At $500\,\text{nm}$, the refractive index contrast between the high-index and the low-index layer can be up to $0.4$, but the absorption length is below $10\mu \text{m}$ even for $10\%$BC/PS, which corresponds to tens of doublelayers. At this wavelength, the absorption length is almost always lower than the scattering length and absorption is dominating. Therefore, no reflectance higher than $0.65$ can be achieved, even though the refractive index contrast is not small.
At $550\,\text{nm}$, the absorption length and scattering length is comparable. So, the final reflectance is limited slightly by the absorption and $100\%$ reflectance can not be achieved.
While at $600\,\text{nm}$, the scattering length is much lower than the absorption length and the scattering dominates. When the thickness of the multilayer is higher than the scattering length, but lower than the absorption length, high or even $100\%$ reflectance is possible. 

\subsection{Full extended gamut}

To explore the full gamut, we varied the ratio of BC and PS in the spheres, the average center-to-center spacing, $s$, sphere diameter, $d$, as well as the sample thickness, $t$. We did not systematically sample this high-dimensionality space, but explored some interesting limiting cases.

\begin{figure}
    \centering
    \includegraphics[width=0.45\textwidth]{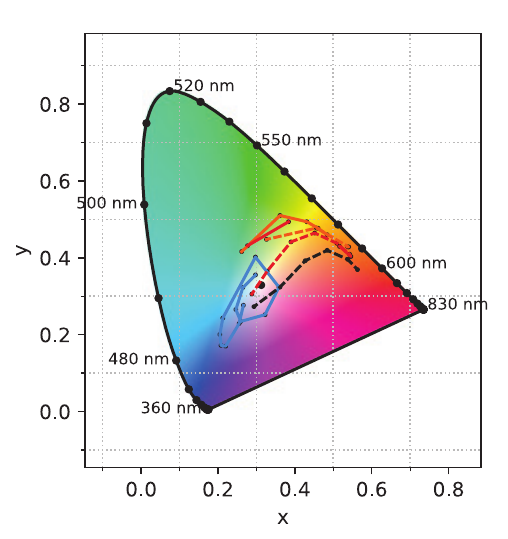}

    \begin{tabular}{||c c c c c||} 
 \hline
 Line style & \% BC & $s/d$ & $s$ [nm] & $t$ [$\mu$m] \\ [0.5ex] 
 \hline\hline
 solid blue & 0 & 1 & 140-390 & 3\\ 
 \hline
 solid orange & 33 & 1 & 300-370 & 3\\
 \hline
 dashed orange & 33 & 2 & 300-440 & 10 \\
 \hline
 solid red & 67 & 2 & 300-370 & 3\\
 \hline
 dashed red & 67 & 2 & 240-480 & 10 \\ 
 \hline
 dashed black & 100 & 2 & 240-390 & 10 \\  
 \hline
\end{tabular}

    \caption{The CIE chromaticity diagram with all the simulated colors from photonic glasses assemblies with different scatterer properties, pigment loadings, and thicknesses. Dots on each line represent colors from photonic glasses with varying sphere diameter, $d$, and spacing, $s$, but with the same pigment loading, thickness, $t$, and diameter/spacing ratio. }
    \label{fig:fullcp}
\end{figure}

\end{document}